\documentclass[reprint,superscriptaddress,showpacs,preprintnumbers,amsmath,amssymb,aps,floats,prl]{revtex4-1}

\usepackage{graphicx}
\usepackage{dcolumn}
\usepackage{bm}

\newcommand{\lra}{\longrightarrow}

\begin{document}

\title{L-valley Electron Spin Dynamics in GaAs}

\author{T. Zhang}
\author{P. Barate}
\author{C.T. Nguyen}
\author{A. Balocchi}
\author{T. Amand}
\author{P. Renucci}
\author{H. Carrere}
\author{B. Urbaszek}
\affiliation{Universit\'e de Toulouse, INSA-CNRS-UPS, LPCNO, 135 avenue de Rangueil, 31077 Toulouse, France\\}
\author{X. Marie}
\email{marie@insa-toulouse.fr}
\affiliation{Universit\'e de Toulouse, INSA-CNRS-UPS, LPCNO, 135 avenue de Rangueil, 31077 Toulouse, France\\}

\date{\today}

\begin{abstract}
Optical orientation experiments have been performed in GaAs epilayers with photoexcitation energies in the 3 eV region yielding the photogeneration of spin-polarized electrons in the satellite L valley. We demonstrate that a significant fraction of the electron spin memory can be conserved when the electron is scattered from the L to the $\Gamma$ valley following an energy relaxation of several hundreds of meV. Combining these high energy photo-excitation experiments with time-resolved photoluminescence spectroscopy of $\Gamma$ valley spin-polarized photogenerated electrons allows us to deduce a typical L valley electron spin relaxation time of 200 fs, in agreement with theoretical calculations.
\end{abstract}

\maketitle
The electron spin dynamics have been studied in great detail for about 50 years in semiconductors thanks to the optical orientation technique \cite{Lampel,Dyakonov}. However all these experiments were performed with optical excitation energies close to the band gap (typically 1.5 - 2 eV in GaAs), yielding the photogeneration of spin-polarized electrons in the $\Gamma$ valley.
In addition to its fundamental aspect, the knowledge of the electron spin dynamics of electrons in the upper valleys is of great interest for spintronic devices such as Spin-LEDs or Spin-Lasers, where the electrical spin injection can lead to electrons populating not only the $\Gamma$ valley but also the satellite $L$ and $X$ valleys whose spin dynamics is almost unknown. In Spin-LEDs based on a Ferromagnetic (FM) layer and Schottky barrier, it was predicted with Monte Carlo simulations that strong electric fields at the interfaces between the FM and the semiconductor layer lead to the redistribution of electrons among several valleys ($L$ and $X$), where the spin relaxation times have been predicted to be much shorter than the one in the $\Gamma$ valley \cite{Saikin,Barry,Wu2012,Mallory}. These upper valley electrons thus play a crucial role in the operation of Spin-LED and Spin laser devices \cite{Mallory}. For the demonstration of the Spin Gunn effect predicted a few years ago it is also essential to get information about the spin relaxation times of high energy electrons in the $L$ valley \cite{Flatte}.
Very little is known about the spin polarization and the spin dynamics of the high energy electrons in these L valleys in GaAs, though intervalley scattering in zincblende semiconductors has for a long time been a subject of theoretical and experimental interest \cite{Gunn,Ridley,Mirlin}. The Dresselhaus intrinsic spin splitting, which is a key parameter for the spin polarization properties has been mainly studied in the close vicinity near the Brillouin zone center $k_0=\Gamma$ \cite{Wu}. The spin-orbit coupling parameters in the upper valleys, for $k_0$=$L$ or $k_0$=$X$ have been calculated recently by different groups \cite{Zunger,Jancu,Wu}. Compared to the $\Gamma$ valley of III-V semiconductors, larger $k$-dependent spin splittings in the surrounding of the L point were predicted \cite{Jancu}. As a consequence the D'Yakonov-Perel spin relaxation mechanism in the $L$ valleys is expected to be very efficient. Multivalley spin relaxation in the presence of high in-plane electric fields in n-type GaAs was investigated theoretically by means of a kinetic spin Bloch equation approach \cite{Wu2012}. This work emphasized the key role played by the spin dynamics of the L electrons in addition to the well known dynamics of $\Gamma$ electrons.

The few experimental investigations of the $L$-valley electron spin polarization were performed by photo-emission spectroscopy in GaAs \cite{Pierce,Drouhin}. In these experiments the GaAs surface is treated with cesium and oxygen to obtain a negative electron affinity. The spin polarization of electrons photoemitted from (110) GaAs following the excitation with circularly polarized light is then measured by Mott scattering. These experiments performed with excitation energy of $\sim$3 eV, leading to the photogeneration of $L$ valley electrons, demonstrated a spin polarization of photo-ionized electrons of about 8\% at low temperature. However since the kinetic energy of electrons was not measured simultaneously with their spin it is difficult to assign this polarization to L electrons only or to a mixture of $L$ and $\Gamma$ electrons. Moreover some depolarization can occur when the electrons photoemitted from GaAs pass through the Cs-O layer; as a consequence $L$ valley electron spin relaxation times are difficult to extract from these photo-emission experiments \cite{Drouhin}.

\begin{figure}
  \includegraphics[width=8.6cm]{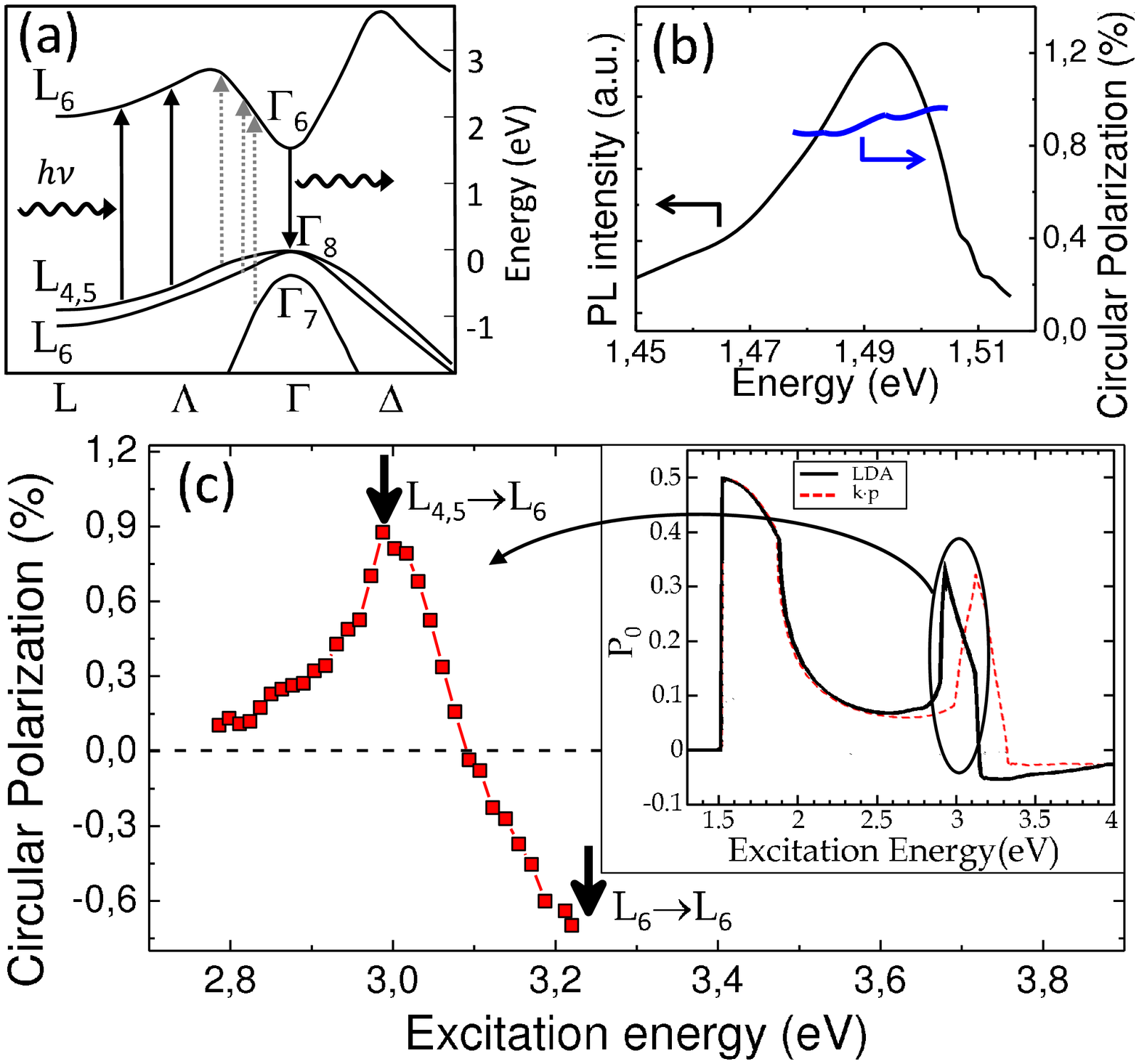}
	\caption{\label{fig:fig1}(a) Schematics of GaAs band structure; the arrows present the optical excitation and detection energies used in (b) and (c); (b) Time-integrated photoluminescence spectrum and the corresponding circular polarization following a $\sigma^+$ polarized laser excitation at an energy $E_{exc}$=2.987 eV; (c) PL circular polarization as a function of the excitation energy. The vertical arrows indicate the energy position of $L$ valley transitions; inset : dependence of the calculated photogenerated electron spin polarization after \cite{Sipe}.}
\end{figure}

In order to get information about the spin polarization of $L$ valley electrons we have performed photoluminescence based optical orientation experiments with laser excitation energies in the range $h\nu$=2.8 to 3.2 eV. 
We have measured the variation of the luminescence polarization detected at the fundamental gap transition ($E_g\sim$1.5 eV) as a function of $h\nu$, Fig. \ref{fig:fig1}(a). We demonstrate that spin polarized electrons photogenerated in the $L$ valley can preserve a significant spin polarization once they have relaxed in the bottom of the $\Gamma$ valley. These experiments combined to classical time-resolved optical orientation experiments performed with excitation energy around 1.5 eV allowed us to mesure a typical $L$ valley electron spin relaxation time $\tau^L_S$ = 200 fs, in good agreement with theoretical predictions.

The investigated sample has been grown by molecular beam epitaxy on nominally undoped (001) GaAs substrates. It consist of 1 $\mu$m Beryllium doped GaAs epilayer with $p_0=10^{18}$ cm$^{-3}$. We present in this Letter the experimental results obtained at 10 K. The excitation source is a mode-locked frequency doubled Ti:Sa laser with a 1.5 ps pulse width and a repetition frequency of 80 MHz. The laser beam propagating perpendicular to the sample surface is focused onto the sample to a 100 $\mu$m diameter spot with an average power $P_{exc}$= 15 mW and its helicity is modulated $\sigma^+/\sigma^-$ with a photo-elastic modulator at a frequency of 50 kHz ; in addition to an increased measurement accuracy this avoids the build-up of the dynamic nuclear polarization \cite{OpticalOrientation}. For the polarized excitation of photoluminescence (PLE) experiments, the time-integrated PL intensity is dispersed by a spectrometer and detected by a Silicon photodiode with a double-modulation lock-in technique. For near band gap excitations, the time-resolved PL measurements are performed with a S20 photocathode streak camera with an overall time resolution of 8 ps \cite{Balocchi}. The Ti:Sa excitation laser is circularly polarized $\sigma^+$ and the resulting PL circular polarization $P_c$ is calculated as $P_c=(I^+-I^-)/(I^++I^-)$. Here, $I^+$ and $I^-$ are the PL intensity components co-polarized and counter-polarized to the  $\sigma^+$ excitation laser.

Fig. \ref{fig:fig1}(b) presents the time-integrated PL spectrum for an excitation energy $E_{exc}$ = 2.987 eV. The PL peak position ($\sim$1.494 eV) is consistent with the band gap shrinkage induced by the high p-doping \cite{Borghs}.  For this photo-excitation energy, Fig. \ref{fig:fig1}(a) shows that four types of optical transitions are allowed. Three of them (dotted lines) will photogenerate electrons in the conduction band (CB) near the $\Gamma$ point through respectively the heavy-hole band $\rightarrow$ CB $\left( \Gamma_8 \lra \Gamma_6\right)$, light-hole band $\rightarrow$ CB $\left( \Gamma_8 \lra \Gamma_6\right)$ and the spin-orbit split-off band $\rightarrow$ CB $\left( \Gamma_7 \lra \Gamma_6\right)$. In the effective mass approximation, this yields the photogeneration of electrons with kinetic energy of 1310 meV, 830 meV and 800 meV respectively. In addition to these 3 optical transitions leading to the photogeneration of $\Gamma_6$ electrons, a strong absorption occurs due to the allowed $L_{4,5}\lra L_6$ transitions in the vicinity of the $L$ valley. Note that the CB $L$ valley minimum lies 296 meV above the $\Gamma$ one. As depicted in Fig. \ref{fig:fig1}(a) there is a large region in $k$ space where the $L_6$ conduction and $L_{4,5}$ valence bands are parallel; this feature together with the fact that the corresponding masses are larger than the ones in $\Gamma$ make these $L$-valley transitions dominant in this spectral region \cite{Greenaway}. For an excitation energy $\sim$200 meV larger, the absorption peak associated to the $L_{4,5}\lra L_6$ transitions vanishes and is replaced by a second peak with a similar amplitude corresponding to the $L_6\lra L_6$ transitions.

Fig. \ref{fig:fig1}(c) displays the PL circular polarization detected at the fundamental gap ($E_{det}$=1.494 eV ) as a function of the excitation energy. The vertical arrows indicate the absorption peaks associated to the $L_{4,5}\lra L_6$ and $L_6\lra L_6$ transitions respectively \cite{Greenaway}. Remarkably we observe a significant polarization though the photogenerated spin-polarized electrons have experienced a very large energy loss before radiative recombination at the bottom of $\Gamma_6$ valley. Let us remind that the PL circular polarization $P_c$ detected at the fundamental gap tracks directly the electron spin polarization $P_s$: $P_s=2P_c$ according to the well know optical selection rules and the fact that the hole spin relaxation time is of the order of 1 ps or less \cite{Dyakonov}. For $E_{exc}$=2.987 eV, we observe a peak in the PL circular polarization $P_c\sim$0.9\%. This peak coincides unambiguoulsly with the absorption peak corresponding to the $L_{4,5}\lra L_6$ transition \cite{Greenaway} ; its position is also similar to the one observed in photo-emission experiments \cite{Pierce}. This demonstrates that the electrons photogenerated in the $L$ valley preserve a fraction of the initial spin polarization after the scattering in the $\Gamma$ valley and subsequent radiative recombination. 

The detected electron spin polarization depends both on (i) the maximum photogenerated spin polarization $P_0$ linked to the optical selection rules imposed by the symmetry of the carrier wavefunctions and (ii) the ratio between the electron spin relaxation time and electron lifetime. The inset in Fig. \ref{fig:fig1}(c) displays the photogenerated electron spin polarization $P_0$ as a function of the optical excitation energy deduced from pseudo potential band structure calculations based on local density approximation (LDA) or 30 bands $\textbf{k}\cdot\textbf{p}$ calculations \cite{Sipe}. We observe a good qualitative agreement between the excitation energy dependence in the 3 eV region of the measured PL circular polarization and the calculated maximum spin polarization despite the great complexity inherent to the calculation of high energy electron wavefunctions. The energy of the measured circular polarization peak (2.987 eV) is closer to the one calculated with LDA (2.90 eV) than with the $\textbf{k}\cdot\textbf{p}$ method (3.15 eV).

\begin{figure}
	\includegraphics[width=8.6cm]{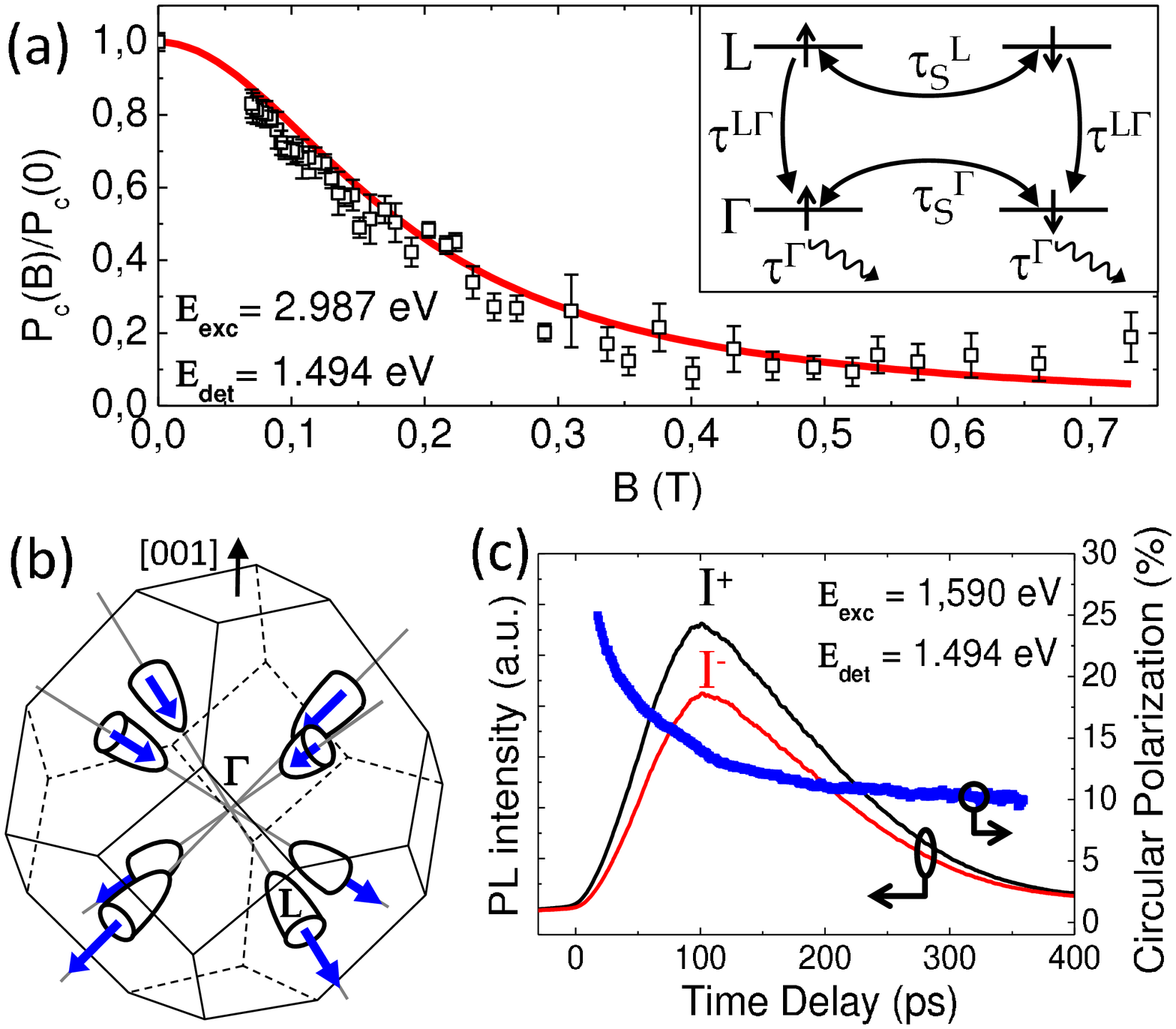}
	\caption{(a) Hanle curve : variation of the PL circular polarization as a function of the transverse magnetic field B. The full line is a Lorentzian curve with $T_S$=140 ps (see text); inset: Schematic representation  of the two-level model including the spin relaxation times in both $L$ and $\Gamma$ valley (see text) (b) Sketch of the Brillouin zone of GaAs displaying the 8 L valleys. The blue arrows represent the photogenerated spins in L valleys (c) Time evolution of the PL circular components $I^+$ and $I^-$ and the corresponding circular polarization $P_c$ for a near band-gap excitation.}
	\label{fig:fig2}
\end{figure}

When the excitation energy increases further we observe in Fig. \ref{fig:fig1}(c) that the measured circular polarization decreases and becomes negative in the 3.2 eV excitation energy region. This is in full agreement with the expected reversed spin polarization when the transition $L_6\lra L_6$ is excited; indeed the spin-orbit splitting energy between the $L_{4,5}$ and $L_6$ bands is $\sim$220 meV \cite{Chadi}. Note that Nastos \textit{et al.} calculated a photogenerated spin polarization in this region $P_0\sim-5\%$ (inset in Fig. \ref{fig:fig1}(c)) \cite{Sipe}. The reversal of the spin polarization sign for the two types of $L$ valley transitions can be explained qualitatively as follows. If we consider the excitation of states in a single $L$ valley by a $\sigma^+$ polarized light propagating along the valley axis (\textit{e.g.} [111]), the photogenerated electron spin polarization would be 100\% for transitions from $L_{4,5}$ to $L_6$ with corresponding wavefunction $\left|(X-iY) \downarrow\right\rangle/\sqrt{2}$ and $\left|S\downarrow\right\rangle$, and -100\% for transitions from $L_6$ to $L_6$ with corresponding wavefunction $\left|(X-iY) \uparrow\right\rangle/\sqrt{2}$ and $\left|S\uparrow\right\rangle$ \cite{Cardona, Pierce}. Taking into account the 8 different $L$ valleys orientations (Fig. \ref{fig:fig2}(b)), the respective joint densities of states for each transitions and a light propagation along the [001] direction (as in the experiments presented here) this yields the calculated value $P_0\sim$30\% for a resonant excitation of the $L_{4,5}\lra L_6$ transition and $P_0\sim-5\%$ for the $L_6\lra L_6$ one (inset in Fig. \ref{fig:fig1}(c)).

Finally let us emphasize that for excitation energies less than 2.8 eV (i.e. smaller than the $L$ valley absorption), we measure in Fig. \ref{fig:fig1}(c) a circular polarization close to zero. The calculations predict in this energy region a maximum photogenerated spin polarization $P_0\sim$10\%. Thus our results indicate that the total contribution of the spin-polarized hot electrons photogenerated in the $\Gamma_6$ conduction band (dotted arrows in Fig. \ref{fig:fig1}(a)) to the PL circular polarization detected at the bottom of the $\Gamma_6$ band is very weak in this excitation energy range. The spin relaxation time for these high energy electrons in the $\Gamma_6$ CB is very short as a result of the large electron $k$ wavector values and the cubic $k$ form of the $\Gamma$ valley Dresselhaus spin-orbit coupling in bulk GaAs \cite{Dyakonov,Wu2012}. As a consequence, the detected $P_c\sim$0.9\% measured for an excitation energy of 2.987 eV can undoubtedly be assigned to the spin-polarized electrons photogenerated in the $L$ valley. This is confirmed by the fact that the contribution of the $L$ valley transitions to the absorption in this energy range is much larger than the $\Gamma$ valley ones\cite{Greenaway,Sipe}. For the sake of simplicity we will neglect in the following the small contribution of these photogenerated $\Gamma_6$ hot electrons.

We have also measured the dependence of the circular polarization on a transverse magnetic field (Voigt configuration) when the spin polarized electrons are photogenerated in the $L$ valley. Fig. \ref{fig:fig2}(a) presents the corresponding Hanle curve for $E_{exc}$=2.987 eV and a detection energy $E_{det}$=1.494 eV. The observed depolarization induced by the magnetic field is another proof that the measured circular polarization of luminescence is the result of optical orientation of electron spins. Because of the fast $L\rightarrow\Gamma$ scattering time, the Hanle curve can be well described  by a simple Lorentzian function which takes into account only the electron spin relaxation time $\tau^\Gamma_S$ and the electron lifetime $\tau^\Gamma$ in the $\Gamma$ valley: $P(B)/P(0)= [1+ (\Omega\cdot T_S)^2]^{-1}$ where $\Omega=g\mu_B B/\hbar$, $g=-0.44$ is the $\Gamma$ electron $g$ factor and  $\mu_B$ the Bohr magneton. The $\Gamma$ electron spin lifetime $T_S$ writes simply $(T_S)^{-1}= (\tau_S^\Gamma)^{-1} +(\tau^\Gamma)^{-1}$. The full line in Fig. \ref{fig:fig2}(a) is the result of a fit with $T_S$=140 ps in satisfactory agreement with the direct measurement by time-resolved photoluminescence spectroscopy presented below.

In order to get some quantitative information on the electron spin dynamics in the $L$ valley from the measured polarization of the luminescence displayed in Fig. \ref{fig:fig1}(c), one has to take into account the optical selection rules in the $L$ valley and the electron spin relaxation time which can occur both in the $\Gamma$ and $L$ valleys.

In contrast to the well-known optical selection rules yielding the photogeneration of spin-polarized electrons in the $\Gamma_6$ valley, the calculated photogenerated spin polarization in the $L$ valley requires to consider the 8 different $\left\langle111\right\rangle$ valleys whose orientation are different from the [001] $\sigma^+$ polarized light propagation axis (see Fig. \ref{fig:fig2}(b)). For the $L_{4,5}\lra L_6$ optical transition, it can be shown that the corresponding  spin polarization in the $L$ valley is $P_0^L$ = 50\% considering a quantization axis along [001] \footnote{Supplementary Information} ; this value is consistent with $P_0\sim$30\% calculated by Nastos \textit{et al.} (inset in Fig. \ref{fig:fig1}(c)) for an optical excitation energy resonant with $L_{4,5}\lra L_6$ but which also includes a weak contribution of the hot photogenerated electrons in the $\Gamma_6$ valley characterized by a smaller spin polarization \cite{Sipe}. The maximum circular polarization of the luminescence detected at the fundamental gap $\left( \Gamma_8 \rightarrow \Gamma_6 \right)$ is thus $P_0^L/2$ =25\% (the factor 2 is due to the transitions involving both heavy holes and light holes \cite{OpticalOrientation}). We emphasize that this \textquotedblleft loss\textquotedblright\; of spin polarization arises from symmetry considerations and not from any spin relaxation mechanisms which have been so far neglected.
We have independently measured the spin relaxation time of the electrons in the $\Gamma$ valley by recording the time and polarization resolved photoluminescence spectrum following a direct photogeneration of $\Gamma$ electrons. Fig. \ref{fig:fig2}(c) presents the time evolution of the luminescence co-polarized $I^+$ and counter-polarized $I^-$ to the $\sigma^+$ excitation laser; the excitation energy is $E_{exc}$= 1.590 eV yielding the photogeneration of spin-polarized electrons in the $\Gamma_6$ conduction band only. The measured initial circular polarization of luminescence is $\sim$25\%, in very good agreement with the optical selection rules in bulk GaAs \cite{OpticalOrientation}. From these kinetics we measure $\tau_S^\Gamma \sim$200 ps and $\tau^\Gamma \sim$105 ps. These values are consistent with previous measurements performed in p-doped GaAs epilayers with similar doping values where it was demonstrated that the spin relaxation of thermalized electrons in the $\Gamma$ valley is due to the Bir-Aronov-Pikus mechanism \cite{Zerrouati, Garbuzov, Aronov, Rosen, Wu,OpticalOrientation}.

Finally we have interpreted the experimental results of Fig. \ref{fig:fig1}(c) in the framework of the following simple two levels rate equation system :
\begin{equation}
\left\{
\begin{array}{rcl}
	\frac{dn^L_{+(-)}}{t}&=&\frac{n_{+(-)}^L-n_{-(+)}^L}{2\tau_S^L}-\frac{n_{+(-)}^L}{\tau^{L\Gamma}} \\
	\frac{dn^\Gamma_{+(-)}}{t}&=&\frac{n_{+(-)}^\Gamma-n_{-(+)}^\Gamma}{2\tau_S^\Gamma}-\frac{n_{+(-)}^\Gamma}{\tau^{\Gamma}}+\frac{n_{+(-)}^L}{\tau^{L\Gamma}}
\end{array}\right.
\end{equation}
where $n^L_{+(-)} \left( n^\Gamma_{+(-)} \right)$ is the density of electrons with spins up and down in the $L$ and $\Gamma$ valley respectively, $\tau_S^L$ and $\tau_S^\Gamma$ the electron spin relaxation times in the $L$ and $\Gamma$ valley ; $\tau^{L\Gamma}$ is the $L\rightarrow\Gamma$ relaxation time and $\tau^\Gamma$ the electron lifetime in $\Gamma$ (inset in Fig. \ref{fig:fig2}(a)). The resolution of equations (1) in steady state conditions leads to the calculated circular polarization of the photoluminescence detected on the fundamental gap following a photogeneration of electrons in the $L$ valley :
\begin{equation}
	P_c=\frac{P_0^L}{2}\frac{1}{\left(1+\frac{\tau^\Gamma}{\tau_S^\Gamma}\right)\left(1+\frac{\tau^{L\Gamma}}{\tau_S^L}\right)}
\end{equation}
For an excitation energy $E_{exc}$=2.987 eV yielding the photogeneration of $L_6$ electrons, Fig. \ref{fig:fig1}(c) shows that the measured PL circular polarization is $P_c$=0.9\% and the calculated photogenerated electron spin polarization $P_0^L$= 30\% (inset in Fig. \ref{fig:fig1}(c)). Assuming a $L\rightarrow\Gamma$ transfer time $\tau^{L\Gamma}$=2 ps as measured by ultrafast spectroscopy \cite{Shah,Stanton,Comment}, we deduce from equation (2) that the electron spin relaxation time in the $L$ valley is $\tau_S^L$=200 fs. Following the same procedure we found a very similar spin relaxation time ($\sim$180 fs) in a second sample characterized by a smaller p-doping (not shown). These measured values are in quite good agreement with recent calculations predicting a spin relaxation time $\tau_S^L \sim$ 100 fs in GaAs at room temperature as a result of the strong spin orbit splitting of conduction electrons in the $L$ valley \cite{Wu2012}. As expected the $L$ valley spin lifetime in GaAs is much shorter than the $L$ valley electron spin relaxation time in centro-symmetric materials with weaker spin-orbit interaction such as Silicon or Germanium \cite{Pezzoli}.

In conclusion we have measured the electron spin relaxation time in the satellite $L$ valley in GaAs. In addition to the well documented spin dynamics in the $\Gamma$ valley, it is a key parameter to develop and optimize spin-optronic and spin transport devices where spin-polarized electrons are redistributed among several valleys. Our measured $L$ electron spin relaxation time ($\sim$200 fs) looks at first sight shorter than what is required ($\sim$3 ps) for the spontaneous spin amplification to appear under electric field where the charge Gunn effect appears \cite{Flatte}. Nevertheless we believe that the enhanced electron-electron scattering in highly n-doped GaAs could yield longer spin relaxation times compared to the ones we have measured here and this could allow the experimental demonstration of Spin Gunn effect.

\begin{acknowledgments}
We thank professor M.W. Wu for very useful discussion. We are grateful to C. Fontaine and A. Arnoult from LAAS-CNRS for the MBE growth of the samples.
\end{acknowledgments}

\providecommand{\noopsort}[1]{}\providecommand{\singleletter}[1]{#1}%


\begin{thebibliography}{31}
\bibitem [{Com()}]{Comment}%
  \BibitemOpen
  \href@noop {} {}\bibinfo {note} {Note that the $\Gamma\lra L$ transfer time
  is $\sim$100 fs, much shorter than the $L\lra\Gamma$ one because of the
  smaller density of states in the $\Gamma$ valley linked to the very different
  effective electron masses : $m_\Gamma = 0.067m_0$ and $m_L = 0.23m_0$, with
  $m_0$ representing the free electron mass.}\BibitemShut {Stop}%
\end{thebibliography}
\end{document}